\documentclass[preprint,showpacs,preprintnumbers,amsmath,amssymb,nofootinbib]{revtex4-1}

\usepackage{graphicx}
\usepackage{amsmath}
\usepackage{amssymb}
\usepackage{xcolor}
\usepackage{dcolumn}
\usepackage{epstopdf}

\begin{document}

\title{Constraining fast radio burst progenitors\\ with gravitational lensing
 }

\renewcommand{\thefootnote}{\fnsymbol{footnote}}

\author{Chun-Yu Li$^{1*}$ and Li-Xin Li$^{2}$}
\affiliation{$^{1}$Department of Astronomy, Peking University,\\ Beijing 100871, China;\\
$^{2}$Kavli Institute for Astronomy and Astrophysics, Peking University,\\ Beijing 100871, China}

\footnotetext{Corresponding author(email: licy08@pku.edu.cn)}

\date{\today}

\begin{abstract}
Fast Radio Bursts (FRBs) are new transient radio sources discovered recently. Because of the angular resolution restriction in radio surveys, no optical counter part has been identified yet so it is hard to determine the progenitor of FRBs. In this paper we propose to use radio lensing survey to constrain FRB progenitors. We show that, different types of progenitors lead to different probabilities for a FRB to be gravitationally lensed by dark matter halos in foreground galaxies, since different type progenitors result in different redshift distributions of FRBs. For example, the redshift distribution of FRBs arising from double stars shifts toward lower redshift than of the FRBs arising from single stars, because double stars and single stars have different evolution timescales. With detailed calculations, we predict that the FRB sample size for producing one lensing event varies significantly for different FRB progenitor models. We argue that this fact can be used to distinguish different FRB models and also discuss the practical possibility of using lensing observation in radio surveys to constrain FRB progenitors.

\vspace{0.3cm}

\noindent
{\bf Star formation, Radio sources, Gravitational lenses and luminous arcs, Cosmology}

\vspace{0.3cm}

\end{abstract}

\maketitle

\section{Introduction}
Fast Radio Bursts (FRBs) were discovered in recent years \cite{Thorn2013}. A burst rate of about ten thousand bursts per day over the entire sky was deduced \cite{cor2013}. However, it is still not possible to determine the astrophysical origins and the redshift distribution of FRBs, because no optical counterpart has been discovered yet and hence no redshift of FRBs has been measured. In this paper we propose that strong gravitational lensing (SGL) of FRBs by dark matter halos in foreground galaxies may have a critical role in the study of FRB progenitors.

Radio surveys for FRBs suffer from the restriction in angular resolutions, which makes it difficult to identify the respective optical counterparts. But for transient sources like FRBs, the SGL properties may be used for identifying the progenitor of FRBs. Different FRB progenitor models result in different FRB redshift distributions, and hence different FRB lensing probabilities, that is, different expected sample size for producing one lensing case.

Due to the transient nature of FRBs and the time-delay between images in gravitational lensing, it is possible to identify lensing events of FRBs in a radio survey despite the restriction of angular resolution. The standard way for identifying gravitational lensing images is based on two criteria: (1) their light curves have identical shapes and identical durations; (2) their spectra have the same shape.

The second criteria is critical, because the first criteria is relatively easy to check. If a radio telescope has several frequency channels, then the ratio between fluxes in any two channels can be used to judge if the radio spectra of the two FRB images have the same shape. A similar dispersion measure (DM) may be another criteria to justify a lensing case.

The statistical lensing probability obtained from a long period FRB survey for multiple lensing images may be used to constrain the associated distribution profile in the FRB redshift space. In addition, if a FRB strong lensing event is confirmed in a well-studied lens system, the redshift of the FRB may be inferred.

 We calculate the probability for gravitational lensing of FRBs by the singular isothermal sphere (SIS) dark matter halos in foreground galaxies, assuming that the mass function of the halos is given by the Press-Schechter function and the Universe is described by the standard LCDM model. Later, we build four types of FRB models, each corresponding to a FRB progenitor model with different redshift distributions. Lastly, we calculate the expected FRB sample size for producing one lensing case and compare the results for different FRB models.

\section{Gravitational Lensing Produced by Dark Matter Halos in Galaxies}
Herein we present the statistics and properties for a remote source gravitationally lensed by dark matter halos in foreground galaxies. We assume that all halos have a SIS mass profile in the mass range $10^{10} h^{-1} M_\odot < M < 2\times 10^{13} h^{-1}M_\odot $, where $h$ is the Hubble constant in units of 100 km s$^{-1}$ Mpc$^{-1}$. This mass range roughly corresponds to dark matter halos residing in galaxies. Halos with masses out of this range often have a NFW-type mass profile \cite{nava97} and make only small contribution to the total lensing probability \cite{li02,li03}.

As in \citet{li02}, we assume that the number density of dark matter halos distributed
in mass is described by the Press-Schechter function \cite{pre74}. We
compute the CDM power spectrum using the fitting formula provided by \citet{eis99}. To be consistent with the recent observations of
{\em WMAP9} \cite{hin13}, we assume that the Hubble constant $h = 0.7$, the primordial spectrum index $n_s = 0.97$, the cosmological constant fraction $\Omega_{\Lambda} = 0.72$, and the standard deviation for the primordial density fluctuation $\sigma_8 = 0.82$.

According to the theory of gravitational lensing \cite{sch92}, the probability for a remote point source at redshift $z_S$ lensed by foreground dark matter
halos is given by:
\begin{eqnarray}
	P = \int_0^{z_S} dz_L\,\frac{d D_p}{dz_L}
		\int_0^\infty d M\, n(M,z_L) \sigma(M,z_L) \;,
	\label{ip}
\end{eqnarray}
where $D_p$ is the proper distance
from the observer to a lens at redshift $z_L$, $n(M,z_L) dM$ is the
proper number density of lens objects of masses between $M$ and $M+
dM$, $\sigma(M,z_L)$ is the lensing cross-section of a dark halo of mass
$M$ at $z_L$.

Assuming that $D_L^A$, $D_S^A$, and $D_{LS}^A$ are the angular-diameter distances from the observer to the lens, from the observer to the source, and from the lens to the source, respectively. Then the cross-section for a SIS lens to produce two images with a brightness ratio $ < r$ is \cite{li02}:
\begin{eqnarray}
	\sigma(< r) = \pi \xi_0^2 \left(\frac{r-1}{r+1}\right)^2, \hspace{1 cm}
    \xi_0 = 4\pi\left(\frac{\sigma_v}{c}\right)^2\,
          \frac{D^A_L D^A_{LS}}{D^A_S}\,,
	\label{siga}
\end{eqnarray}
where $\sigma_v$ is the velocity dispersion parameter.

The comoving number density of dark halos formed by redshift $z$ with mass in the range $(M,M+dM)$ is given by:
\begin{eqnarray}
     n(M,z)\,dM = \frac{\rho_0}{M}\, f(M,z)\, dM\,,
\end{eqnarray}
where $\rho_0\equiv\Omega_m\,\rho_{{\rm crit},0}$ is the present mean mass
density in the universe, $f(M,z)$ is the Press-Schechter function \cite{pre74}. The detailed steps for calculating the Press-Schechter function is described in \cite{li02}.

With the above formulas we have calculated the lensing probability for a source object at various redshifts. The results are shown in Figs. 1-4.

\begin{figure}
\begin{center}\includegraphics[angle=0,scale=0.65]{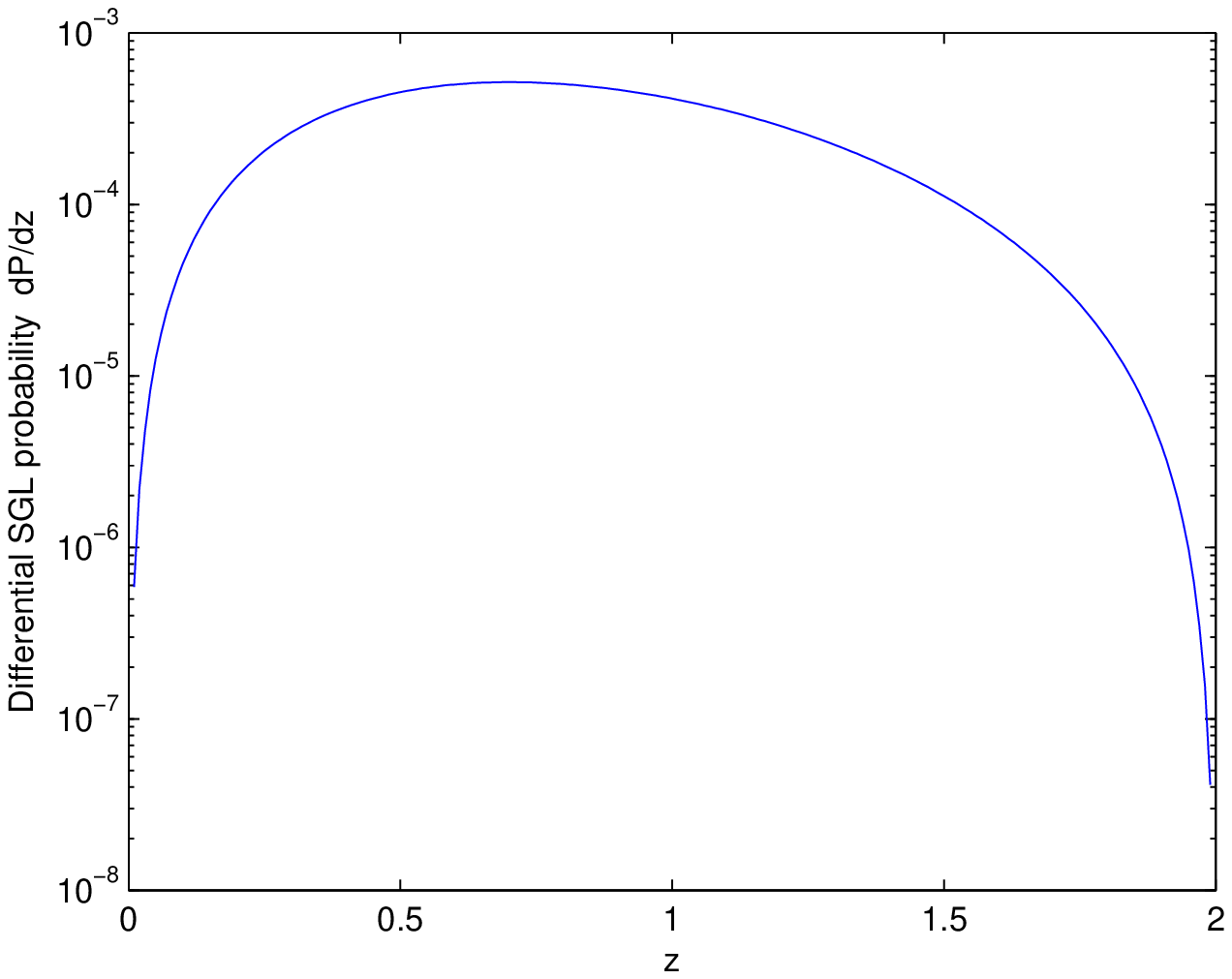}\end{center}
\caption{The differential lensing probability $dP/dz$ for a source at $z_S = 2$. The brightness ratio between the two images is $r \leq 5$.
\label{fig1}}
\end{figure}

\begin{figure}
\begin{center}\includegraphics[angle=0,scale=0.65]{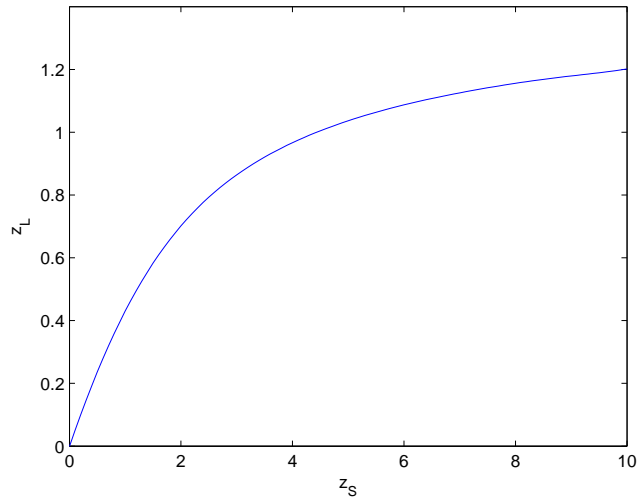}\end{center}
\caption{The lens redshift $z_L$ to produce the maximal differential lensing probability as a function of the source redshift $z_S$.
\label{fig2}}
\end{figure}

Figure \ref{fig1} shows the differential lensing probability for a source located at $z=2$ where the brightness ratio between the two images is assumed to be $r\leq 5$. We see that the source is most likely to be lensed by a halo at $z \approx 0.7 $. Figure \ref{fig2} shows the lens redshift $z_L$ to produce the maximal differential lensing probability as a function of the source redshift $z_S$.

\begin{figure}
\begin{center}\includegraphics[angle=0,scale=0.65]{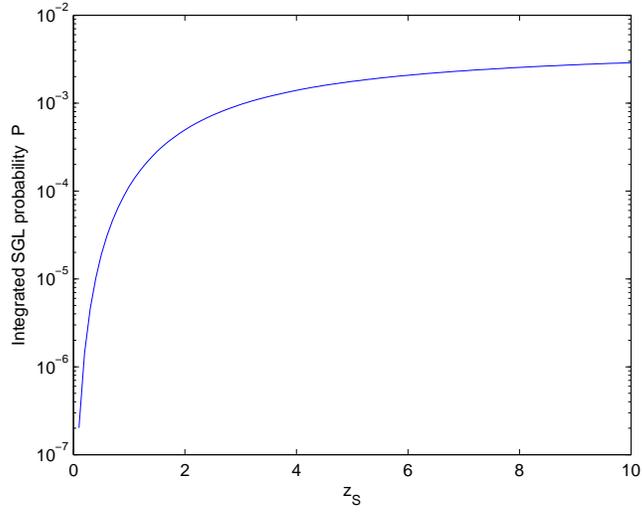}\end{center}
\caption{The integrated lensing probability $P$ as a function of the source redshift, for brightness ratio $r \leq 5$.
\label{fig3}}
\end{figure}

Figure \ref{fig3} shows the integrated lensing probability as a function of the source redshift. The integrated probability increases with the redshift of the source. As the redshift goes beyond 10, the curve becomes somewhat flat which is caused by the fact that the proper distance becomes insensitive to the variation of redshift at very large redshift.

\begin{figure}
\begin{center}\includegraphics[angle=0,scale=0.65]{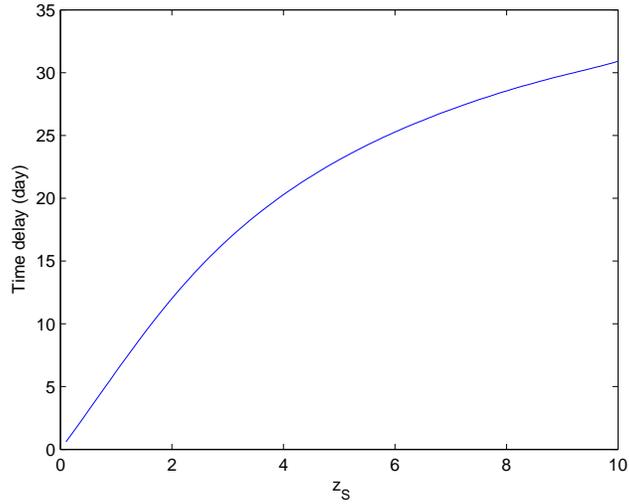}\end{center}
\caption{The typical time delay between the two images as a function of the source redshift $z_S$, for a lens halo of mass $M=10^{12} M_\odot h^{-1}$ at redshift $z_L$ determined in Figure. \ref{fig2}.
\label{fig4}}
\end{figure}

In Fig. \ref{fig4}, the typical time delay between the two lensing images is shown as a function of the source redshift $z_S$, for a lens halo of mass $M=10^{12} M_\odot h^{-1}$ at $z_L$ determined in Fig. \ref{fig2}.

\section{FRB Progenitor Models and FRB redshift Distribution}
Since the discovery of FRBs, several models
have been proposed for interpreting their nature. These models include delayed collapse of supra-massive
neutron stars to black holes \cite{falcke13}, special magnetar radio flares \cite{popov07,popov13},
mergers of double neutron stars \cite{totani13}, mergers of binary white dwarfs \cite{kashiyama13}, flaring stars \cite{loeb13}, and a small fraction of events associated with gamma-ray bursts \cite{zhang14,deng14}.

Here, we divide all the proposed models into two types according to their origins: single star models, and double star models. Accordingly, we build two models for the redshift distribution of FRBs: the single star model, and the double star model. The single star model simply follows the star formation rate ($SFR$). In contrast, the double-star model follows the SFR with an evolving time delay \cite{virg11}. For both models, we first calculate their comoving number density as a function of redshift, then calculate their redshift distribution by including the comoving volume element and the cosmological time dilation.

 For the single star model, the life time of the progenitor can be neglected compared to the cosmological time. The comoving number density rate of the source is then given by
the $SFR$ fitted with the formula  \cite{col01}:
\begin{eqnarray}
	\rho_0(z)= SFR(z) = \frac{a+b z}{1+(z/c)^d} \;. \label{cole}
\end{eqnarray}
where $(a,b,c,d) = (0.0157, 0.118, 3.23, 4.66)$  \cite{li08}.

For the double star model,
we insert the merger time delay into $SFR$ to get the merger rate $\rho_i(z)$, where $i=1, 3, 5$, corresponding to three different evolution timescales: $\Delta t(i) = 1, 3, 5 \ Gyr$ respectively \cite{virg11}
 \begin{eqnarray}
 \rho_i(z)=SFR(z'), \ \  z'=T^{-1}(T(z)+\Delta t(i)),
\end{eqnarray}
 where $T(z)$ is the cosmological look-back time given by the following equation
\begin{eqnarray}
  T(z)=\int_{0}^{z}t(z')dz' =
{\int_{0}^{z} \frac{1}{H_{\rm 0}} \frac{1}{(1+z')
(\Omega_m(1+z')^3+\Omega_\Lambda)^{0.5}}}dz',
\end{eqnarray}
and $T^{-1}$ is the inverse of the function $T(z)$.

\begin{figure}
\begin{center}\includegraphics[angle=0,scale=0.65]{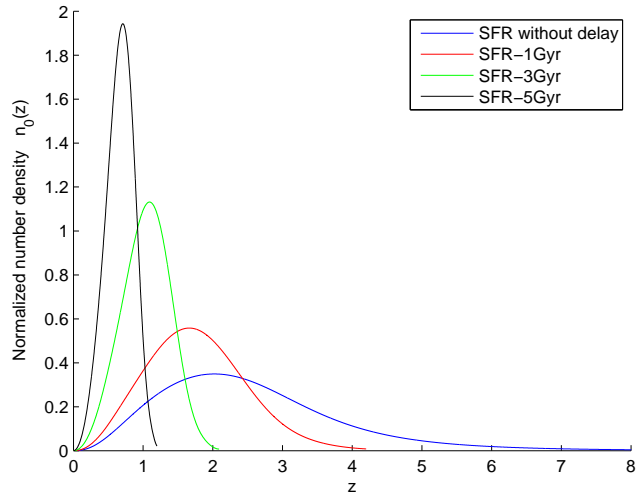}\end{center}
\caption{The normalized number density distribution in the redshift space for different FRB progenitor models, calculated with equation (\ref{f_grb}) and the integration over $z$ being normalized to 1.
\label{fig5}}
\end{figure}

Taking into account the comoving volume element and the cosmological time dilation, the number density of FRBs in the redshift space and per unit time in the observer frame is given as:
\begin{eqnarray}
	n_0(z) \equiv \frac{\rho_i(z)}{1+z}\frac{dV_{com}}{d z} \;,	
	\label{f_grb}
\end{eqnarray}
where $V_{com}$ is the comoving volume.
 Figure \ref{fig5} shows the redshift
distribution calculated from equation (\ref{f_grb}) for different values of the time delay $\Delta t$.

 The observable number density $n(z)$ differs from the $n_0(z)$ because of the observational selection effect. Assuming that FRBs have an intrinsic luminosity function described by a single power-law with a redshift evolution factor:
$\phi(L,z)\propto (L/L_0(z))^{-\beta}$, where $L_0(z) \propto (1+z)^{\alpha}$. The observable number rate of FRBs between $(z, z+dz)$ is then:
\begin{eqnarray}
   dN(z)=K \ n_0(z)dz\int_{L_{min}(z)}^\infty \phi(L)dL, \label{sel}
\end{eqnarray}
where $K$ is a constant determined by detection efficiency. $L_{min}(z)$ is the lower limit of luminosity caused by the telescope detection flux limit $F_0$
\begin{eqnarray}
   L_{min}(z)=4\pi F_0 {D_L(z)}^2,\end{eqnarray} where $D_L(z)$ is the luminosity distance to the source.
Thus, the normalized observable FRBs number density rate is
\begin{eqnarray}
   n(z)=\frac{dN{(z)/dz}}{\int_{z=0}^{\infty}dN(z)}=\frac{{D_L(z)}^{-2(\beta-1)}n_0(z)(1+z)^{\alpha(\beta-1)}}{\int_{z=0}^{\infty}{D_L(z)}^{-2(\beta-1)}n_0(z)(1+z)^{\alpha(\beta-1)}dz} .
   \label{n_z}
\end{eqnarray}

\begin{figure}
\begin{center}\includegraphics[angle=0,scale=0.65]{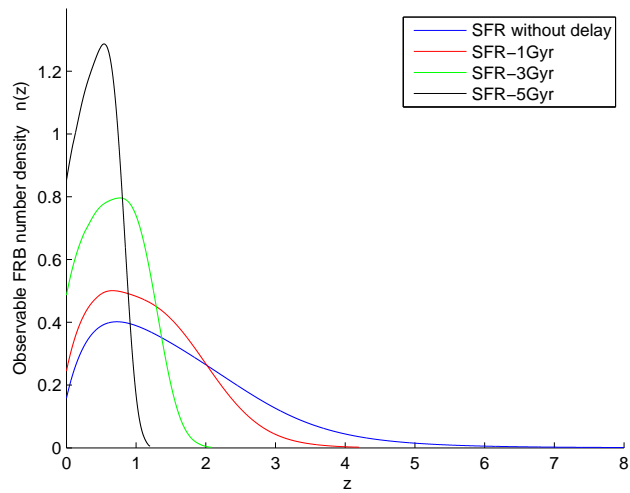}\end{center}
\caption{The normalized observable number density distribution for different FRB progenitor models including selection effect, calculated with equation (\ref{n_z}).
\label{fig6}}
\end{figure}

 Following \cite{Tan13} and \cite{Gus03}, we adopt $\alpha=\beta=2$. Figure \ref{fig6} shows the redshift distribution of the observable number density $n(z)$ calculated from equation (\ref{n_z}) for different values of the time delay $\Delta t$.

Both Figs. \ref{fig5} and \ref{fig6} indicate that as the evolution timescale increases, the peak of the FRB distribution moves toward a lower redshift.

\section{Lensing Probability for Different FRB models}
 To compare the lensing probability prediction with observations we must also consider the effect of magnification bias. According to \citet{tur84}, if the sample of source objects has a power-law flux distribution $N \propto f^{-\beta} $, the
magnification bias factor $B$ by which lensed objects will be overrepresented in any particular observed sample is given by:
\begin{eqnarray}
    B = \frac{2}{3-\beta}\,\mu_m^{\beta -1}\,,
    \label{bias2}
\end{eqnarray}
where $ \mu_m$ is the minimum total magnification of the lensing event.

For a SIS lens the total magnification $\mu$ is related to brightening ratio $r$ by $\mu = 2+ 2/(r-1) $. If we require $r \leq 5$ (a too large value of $r$ may result in that one of the images becomes invisible), we get $\mu_m = 2.5$.

 With the SGL probability $P(z)$ and the FRB number density distribution $n(z)$, we can calculate the expected size of the observation sample for getting one FRB lens pair with $r \leq 5$ by:
\begin{eqnarray}
   N_0 = \frac{1}{\int_{0}^{\infty}B P(z)n(z)dz}.
\end{eqnarray}

Adopting $\alpha = 2$ \cite{Tan13} and $\beta = 2$ \cite{Gus03}, we calculated the value of $N_0$ for different FRB models and show the results in Table \ref{tab1}. We can see that the expected size of the FRB observation sample $N_0$ varies significantly for different FRB progenitor models. This is because that different evolution time scales in different models lead to different redshift distribution of FRBs.

Thus a large FRB lens survey collecting more than $1,000$ FRBs in a complete sample may offer a way to distinguish FRB progenitors.

\section{Discussion and Conclusions}

We have calculated the probability for a FRB gravitationally lensed by dark matter halos in foreground galaxies. For a given population of FRBs, the needed sample size for producing a lensing event depends on the redshift distribution of the FRBs. We have built four different FRB redshift distribution models for four types of FRB progenitors. We found that, for the four different models, the sizes of the FRB samples needed for producing a lensing event are very different. The results are summarized in Table \ref{tab1}.

From Table \ref{tab1} we can note that, for typical parameters adopted in the paper, the number of FRBs contained in the sample can differ by an order of magnitude. At least in principle, we can carry out a large FRB lens survey jointly with many radio telescopes to distinguish different FRB progenitors.

The observation of FRBs has revealed a burst rate of $\sim 10^4$ per day \cite{cor2013}, which is sufficiently high. However, we note that the sample size calculated here is for a continuous survey in a fixed field of the sky. If the survey is not continuous, a lensing event may be missed out due to the transient nature of FRBs: for example, for a double-image lensing event, an image may not be detected if it occurs during a gap between two time segments of a survey because of the time-delay of gravitational lensing. Then this lensing event will not be identified.

For most of the ground based interferometers, the observation time for a fixed field of the sky is discontinuous with gaps. All FRBs occurring during those gaps will miss the detection, including those FRB lensing events. For example, if the observation is carried on for $12$ hours per day, the sampling efficiency for FRB lensing events will be $f = 1/2$ since one of the double images in a lensing event may not be detected. In this case, the expected sample size given in Table \ref{tab1} should be enlarged by a factor $1/f=2$. However, the ratio of the sample size between different models is not affected.

If the fixed field is $S (deg^2)$ and the survey program lasts $T (days)$,  we have:
\begin{eqnarray}
    S*T = \frac{S_0 N}{f^2 R},
    \label{st01}
\end{eqnarray}
where $S_0= 41,253 deg^2$ is the area of the whole sky, $N$ is the total number of FRBs needed in the sample, and $R$ is the rate of FRBs.
Given that $R = 10^4$ per day over the entire sky, $f = 1/2$, and $N= 490 $ for the FRB model without a time delay, we have $S*T = 8,000 deg^2 \cdot day$. This is a typical value of $S*T$ needed for a radio survey for FRB lensing events.

From the above results we can predict that if a continuous FRB survey for a fixed field of sky of approximately $10 {\rm deg}^2$ lasts approximately 2 years, it will be possible to constrain FRB progenitors without counterpart observation.

\section*{Acknowledgments}

The authors thank Bing Zhang for helpful comments and very useful insights. This work was supported by the National Basic Research Program (973
Program) of China (Grant No. 2014CB845800) and the NSFC grants
(No. 11373012).

\clearpage

\clearpage
~
\vspace{8cm}
\begin{table}[h]
\centering
\caption{The expected sample size for getting one lens event for different FRB models\label{tab1}}
\vspace{0.5cm}
\begin{tabular}{lcccc}
\hline
\hline
~Model & SFR & SFR1 & SFR3 & SFR5 \\
\hline
~Parameter  & ~$\Delta t = 0$~ & ~$\Delta t = 1 Gyr$~ & ~$\Delta t = 3 Gyr$~ & ~$\Delta t = 5 Gyr$~ \\
~Expected $N_0$~~~~ & ~ $490$~ & ~$880$~ & ~$2690$~ & ~$8730$~ \\
\hline
\end{tabular}

\end{table}

\end{document}